# Tunable vortex bound states in multiband CsV$_3$Sb$_5$-derived kagome superconductors

*Zihao Huang$^{1,2\#}$, Xianghe Han$^{1,2\#}$, Zhen Zhao$^{1,2}$, Jinjin Liu$^{3,4}$, Pengfei Li$^1$, Hengxin Tan$^5$, Zhiwei Wang$^{3,4}$, Yugui Yao$^{3,4}$, Haitao Yang$^{1,2,6}$, Binghai Yan$^5$, Kun Jiang$^1$, Jiangping Hu$^1$, Ziqiang Wang$^7$, Hui Chen$^{1,2,6*}$, Hong-Jun Gao$^{1,2,6*}$*

[1] Beijing National Center for Condensed Matter Physics and Institute of Physics, Chinese Academy of Sciences, Beijing 100190, China

[2] School of Physical Sciences, University of Chinese Academy of Sciences, Beijing 100190, China

[3] Centre for Quantum Physics, Key Laboratory of Advanced Optoelectronic Quantum Architecture and Measurement (MOE), School of Physics, Beijing Institute of Technology, Beijing 100081, China

[4] Beijing Key Lab of Nanophotonics and Ultrafine Optoelectronic Systems, Beijing Institute of Technology, Beijing 100081, China

[5] Department of Condensed Matter Physics, Weizmann Institute of Science, Rehovot 7610001, Israel

[6] Hefei National Laboratory, Hefei 230088, China

[7] Department of Physics, Boston College, Chestnut Hill, MA 02467, USA

[#] These authors contributed to this work

[*] Corresponding authors.

Email addresses: hchenn04@iphy.ac.cn (H. Chen), hjgao@iphy.ac.cn (H.-J. Gao).

# ABSTRACT

Vortices and bound states offer an effective means of comprehending the electronic properties of superconductors. Recently, surface-dependent vortex core states have been observed in the newly discovered kagome superconductors $CsV_3Sb_5$. Although the spatial distribution of the sharp zero energy conductance peak appears similar to Majorana bound states arising from the superconducting Dirac surface states, its origin remains elusive. In this study, we present observations of tunable vortex bound states (VBSs) in two chemically-doped kagome superconductors $Cs(V_{1-x}Tr_x)_3Sb_5$ (Tr=Ta or Ti), using low-temperature scanning tunneling microscopy/spectroscopy. The $CsV_3Sb_5$-derived kagome superconductors exhibit full-gap-pairing superconductivity accompanied by the absence of long-range charge orders, in contrast to pristine $CsV_3Sb_5$. Zero-energy conductance maps demonstrate a field-driven continuous reorientation transition of the vortex lattice, suggesting multiband superconductivity. The Ta-doped $CsV_3Sb_5$ displays the conventional cross-shaped spatial evolution of Caroli-de Gennes-Matricon bound states, while the Ti-doped $CsV_3Sb_5$ exhibits a sharp, non-split zero-bias conductance peak (ZBCP) that persists over a long distance across the vortex. The spatial evolution of the non-split ZBCP is robust against surface effects and external magnetic field but is related to the doping concentrations. Our study reveals the tunable VBSs in multiband chemically-doped $CsV_3Sb_5$ system and offers fresh insights into previously reported Y-shaped ZBCP in a non-quantum-limit condition at the surface of kagome superconductor.

Key words: Kagome superconductor, chemical doping, vortex lattice transition, vortex bound states, Majorana bound states

## 1. Introduction

Abrikosov vortices, which are topological defects of superconducting order in type-II superconductors, are considered to be exotic quantum objects [1–3]. The vanishing of superconducting order inside the vortex leads to the emergence of in-gap bound states within their core [4]. Understanding the quantum structure of these vortex bound states (VBSs) provides an effective way to comprehend the electronic properties of superconductors, such as the symmetry of superconducting order [5–7] and the topology of electronic states [8]. For instance, as a conventional s-wave superconductor, vortex cores contain Caroli-de Gennes-Matricon (CdGM) states that are spin degenerate and form a quasi-continuum with energy level $E_\mu = \pm \mu \Delta^2/E_F$ ($\mu = 1/2, 3/2, 5/2, \ldots$) [4,9]. However, the quantum structure of VBSs differs with changes in the symmetry of superconducting order. Wang-MacDonald vortex cores have been predicted [10] and observed [11] in a *d*-wave superconductor. Additionally, Majorana zero mode (MZM), a charge-neutral fermion with non-Abelian statistics [12,13], have been predicted to exist in the vortex core of chiral $p_x + ip_y$ superconductor [5,6] or systems that combine topological surface states (TSSs) and superconductivity [14,15]. Abrikosov vortex can also form an ordered vortex lattice due to the inter-vortex repulsion [16]. The morphology of the vortex lattice inherits the symmetry of the underlying electronic structure [17,18] and reflects the details of the pairing mechanism [19]. Therefore, the vortex lattice and VBSs play crucial roles in exploring the properties of superconductivity.

Recently, there has been a surge of interest in the newly-discovered kagome superconductor $AV_3Sb_5$, due to its coexistence of $Z_2$ topology [20], electron nematic order [21–23], unusual symmetry-breaking [24–26] and time reversal symmetry breaking [27,28] charge orders. In addition to its competing charge orders, the superconducting phase in $AV_3Sb_5$ system is equally intriguing. For instance, researchers have observed an unconventional V-shaped superconducting gap [24], anisotropic critical field [29], the signature of multi-band superconductivity [30], and surface-dependent VBSs [31]. Furthermore, two superconducting domes have been reported in the pressured [32–34] and chemical-doped [35,36] phase diagrams. Understanding the nature of superconductivity and its interplay with intertwined electronic orders represents a major frontier in this emerging research field.

In this study, we utilized low-temperature scanning tunneling microscopy/spectroscopy (STM/S) to investigate Ti and Ta-doped kagome superconductors $CsV_3Sb_5$, where we observed two distinct full-gap-pairing superconductivities and tunable VBSs insides Abrisokov vortices. In contrast to the multiple charge orders and V-shaped paring gap observed in pristine $CsV_3Sb_5$, both Ti and Ta-doped $CsV_3Sb_5$ displayed uniform U-shaped superconducting gap paring without long-range charge orders. Increasing the out-of-plane magnetic field led to the reorientation of the vortex lattice with different rotatory angles for Ti and Ta-doped samples. Additionally, in the vortex of Ta-doped sample, we observed the emergence of ordinary CdGM states featuring X-type spatial evolution, while a sharp zero-bias conductance peak (ZBCP) was observed in the vortex cores of Ti-doped $CsV_3Sb_5$, featuring Y-type spatial evolution that persisted non-split over a long distance. The ZBCP with Y-type spatial evolution exist in each vortex which were robust against surface effects and external magnetic fields up to 0.16 T. The decay distance of ZBCP decreased as Ti doping concentrations increased from $x$=0.05 to $x$=0.09.

## 2. Materials and methods

### 2.1. Single crystal growth of the doped $CsV_3Sb_5$ samples

Single crystals of Ti-doped and pristine $CsV_3Sb_5$ single crystals were grown from Cs liquid (purity 99.98%), V powder (purity 99.9%), Ti shot (purity >99.9%) and Sb shot (purity 99.999%) via a modified self-flux method [37]. Single crystals of Ta doped $CsV_3Sb_5$ were grown by the self-flux method [38].

### 2.2. Scanning tunneling microscopy/spectroscopy

The samples used in the STM/S experiments are cleaved at low temperature (13 K) and immediately transferred to an STM chamber. Experiments were performed in an ultrahigh vacuum ($1\times10^{-10}$ mbar) ultra-low temperature STM system equipped with 11 T magnetic field. All the scanning parameters (setpoint voltage and current) of the STM topographic images are listed in the figure captions. The base temperature in the low-temperature STS is 420 mK and the electronic temperature is 620 mK, calibrated using a standard superconductor, Nb crystal. Unless otherwise noted, the d$I$/d$V$ spectra were acquired by a standard lock-in amplifier at a modulation frequency of 973.1 Hz. Non-superconducting tungsten tips were fabricated via electrochemical etching and calibrated on a clean Au(111) surface prepared by repeated cycles of sputtering with argon ions and annealing at 500 °C.

### 2.3. Density-functional theory (DFT) calculations

Calculations are performed within the DFT as implemented in VASP package [39]. The generalized-gradient-approximation as parametrized by Perdew-Burke-Ernzerhof [40] for the exchange-correlation interaction between electrons is employed in all calculations. Zero damping DFT-D3 vdW correction [41] is also employed in all calculations while spin-orbital coupling is not included. The supercell method is used to calculate the formation energy. The supercells with Ti substitutions are fully relaxed until the remaining forces on the atoms are less than 0.005 eV/Å. The k-meshes of 6×6×3 and 3×3×4 are used to sample the Brillouin zones of the 2×2×2 and 3×3×1 supercell of CsV$_3$Sb$_5$, respectively. A cutoff energy of 300 eV for the plane-wave basis set is used.

## 3. Results and discussion

*3.1. Doping-induced suppression of Charge density wave (CDW) and emergence of U-shape superconducting gap of Ti-doped and Ta-doped CsV$_3$Sb$_5$*

Ti-doped and Ta-doped CsV$_3$Sb$_5$ both retain the hexagonal symmetry (space group P 6/mmm) of the parent CsV$_3$Sb$_5$ structure, with Ti/Ta replacing some of the V atoms in the V-Sb kagome layers (Fig. 1a). The STM topography of the Sb surface reveals randomly distributed dark spots for the Ti-doped samples (upper panel of Fig. 1b) but bright spots for Ta-doped samples (lower panel of Fig. 1b). We attribute these additional spots in the STM image of the Sb surface to the chemical dopants in the underlying V-Sb kagome layer. The higher topographic intensity of Ta compared to Ti dopants in the STM image may be attributed to distinct charge transfer effect.

The Sb surfaces of Cs(V$_{1-x}$Ti$_x$)$_3$Sb$_5$ ($x$=0.05) and Cs(V$_{1-x}$Ta$_x$)$_3$Sb$_5$ ($x$=0.14) both exhibit the absence of long-range bi-directional 2$a_0$ and unidirectional 4$a_0$ charge orders (Fig. 1c, d and Fig. S1a, b online), which are present in pristine CsV$_3$Sb$_5$ [24,26]. In addition, distinct from the V-shaped gap paring in pristine compounds (Fig. S2b online), both Ti and Ta doped samples display U-shaped gap pairing in the d$I$/d$V$ linecuts (Fig. 1e, f). The U-shaped gaps in Ti and Ta doped CsV$_3$Sb$_5$ might arise from the shift of van Hove singularities induced by the chemical doping [35,42]. The half peak-to-peak value of the gap is about 0.60 meV for Cs(V$_{1-x}$Ti$_x$)$_3$Sb$_5$ ($x$=0.05) ($T_c$=3.5 K [35]) and 0.77 meV for Cs(V$_{1-x}$Ta$_x$)$_3$Sb$_5$ ($x$=0.14) ($T_c$=5.5 K [42]).

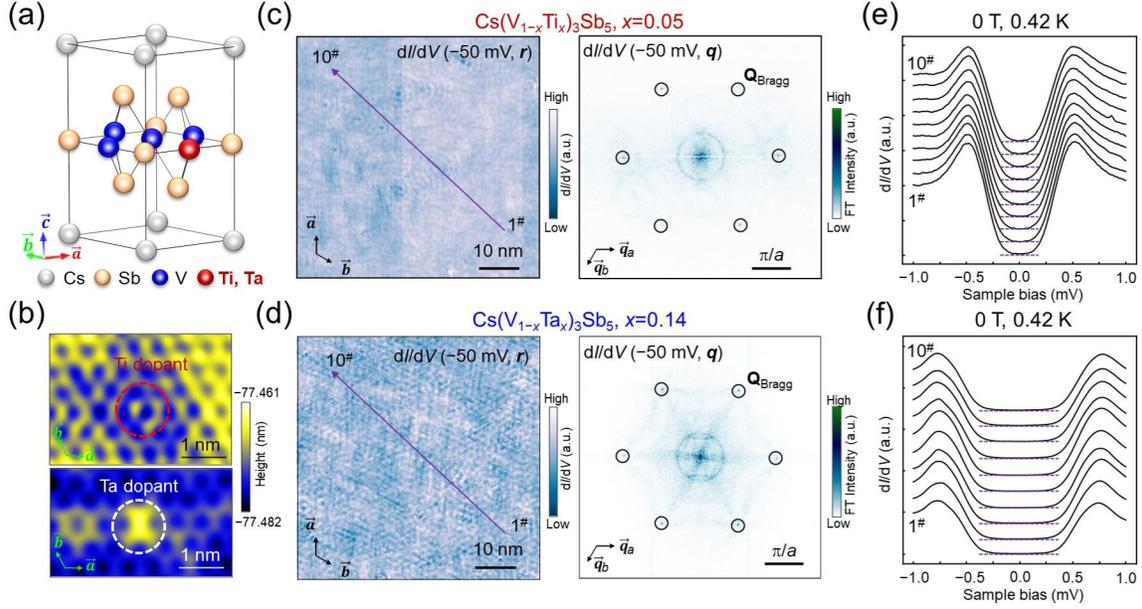

**Fig. 1.** Doping-induced suppression of CDW and emergence of U-shape superconducting gap of Ti-doped and Ta-doped CsV$_3$Sb$_5$. (a) Schematic of atomic structure of doped CsV$_3$Sb$_5$ crystal with Cs atoms. The Ti or Ta atoms (red) replace the V atoms (blue) in Kagome lattice. (b) Atomically-resolved STM images showing the Sb surface of Ti-doped (upper panel) and Ta-doped (lower panel) CsV$_3$Sb$_5$, respectively, showing that the lower density of state around Ti dopants but higher intensity of DOS around Ta dopants (Sample bias: $V$=−100 mV. Setpoint: $I$=2 nA). (c) d$I$/d$V$(−50 mV, $r$) and corresponding Fourier transform d$I$/d$V$(−50 mV, $q$) of the Sb surface Cs(V$_{1-x}$Ti$_x$)$_3$Sb$_5$ ($x$=0.05) obtained at 4.2 K, showing the absence of long-range CDWs ($V$=−50 mV, $I$=1 nA, Modulation voltage: $V_{mod}$=5 mV). (d) d$I$/d$V$(−50 mV, $r$) and d$I$/d$V$(−50 mV, $q$) of the Sb surface Cs(V$_{1-x}$Ta$_x$)$_3$Sb$_5$ ($x$=0.14) obtained at 4.2 K, showing the absence of long-range CDWs as well ($V$=−50 mV, $I$=1 nA, $V_{mod}$=5 mV). (e) The d$I$/d$V$ linecut along the purple arrow in (c), showing the uniform superconducting gap over the region ($V$=−1 mV, $I$=1 nA, $V_{mod}$=0.1 mV). (f) The d$I$/d$V$ linecut along the purple arrow in (d), showing the uniform superconducting gap over the region ($V$=−1 mV, $I$=1 nA, $V_{mod}$=0.1 mV). The spectra in (e-f) are vertically shifted for clarity, and the horizontal dash lines highlight the positions of zero density of states for each curve.

### 3.2. The reorientation of vortex lattice in Ti-doped and Ta-doped CsV$_3$Sb$_5$

To further investigate the superconducting nature, we employ external magnetic fields perpendicular to the sample surfaces ($B_z$) to map the Abrikosov vortices of CsV$_3$Sb$_5$-derived kagome superconductors. At $B_z < H_{c2}$, achieved by zero-field cooling, relatively ordered hexagonal vortex lattice are observed in the zero-energy d$I$/d$V$ maps of both Ti-doped (Fig. 2a-c)) and Ta-doped samples (Fig. 2f-h), indicating weak vortex pinning effects. As $B_z$ increases from 0.0 T to 0.3 T, the lattice constants of the vortices decrease with the magnitude of the field, as expected in type-II superconductors [43]. The density of vortex flux, dependent on $B_z$, yields a single magnetic flux quantum of about $1.99\times10^{-15}$ Wb for the Ti-doped sample (Fig. 2d) and about $2.02\times10^{-15}$ Wb for the Ta-doped sample (Fig. 2i).

In addition to the density of vortices, we observe a rotation of the vortex lattice orientation with increasing $B_z$ (Fig. 2a-c, 2f-h). The angle $\alpha$ is defined as the angle between the nearest inter-vortex direction $\vec{B}$ and the crystalline axis direction $\vec{b}$. The nearest inter-vortex distance $L$ is measured through the three nearest inter-vortex directions. Consequently, the $B_z$-dependent values of $\alpha$ and $L$ (Fig. 2e, j) reveal vortex lattice orientation transitions, which occur differently in Ti- and Ta-doped samples. For the Ti-doped sample, the vortex lattice is aligned with the atomic lattice ($\alpha = 0°$) at low $B_z$ (from 0.05 T to 0.15 T). As the $B_z$ increase to 0.17 T, the orientation of vortex lattice undergoes a continuous change, gradually approaching a stable phase with $\alpha = 30°$ (Fig. 2e). In contrast, for the Ta-doped sample, the vortex lattice orientation begins with $\alpha = 15°$ at low $B_z$ and gradually increases to a stable phase with $\alpha = 30°$ at around 0.12 T (Fig. 2j).

The orientation of the superconducting vortex lattice is primarily influenced by inter-vortex repulsion, vortex pinning, and thermal fluctuations [16,44,45]. In the case of chemically-doped $CsV_3Sb_5$, where the vortex lattice is ordered hexagonally, inter-vortex repulsion dominates. The orientation of the vortex lattice is affected by Fermi surface anisotropy and gap anisotropy as suggested by the non-local corrections of London model [46]. For a hexagonal lattice, it is expected that the vortex lattice aligns itself with the mirror planes of the hexagonal point group, resulting in $\alpha = 0°$ or $30°$ [16,46]. However, in a multiband superconductor like $MgB_2$ [17,47], increasing magnetic field could firstly suppress the small superconducting gap and then leave the superconducting current of vortex dominated by the Cooper pairs with larger gap size. The vortex lattice therefore may change its orientation similar to the case of $MgB_2$ [17,47]. Hence, the observed rotation of the vortex lattice indicates the presence of multiband superconductivity in both doped samples, which is consistent with the multiband superconductivity of pristine $CsV_3Sb_5$ [30]. The different angles of orientation transitions suggest the distinct superconducting order parameters in multiband Ti-doped and Ta-doped $CsV_3Sb_5$.

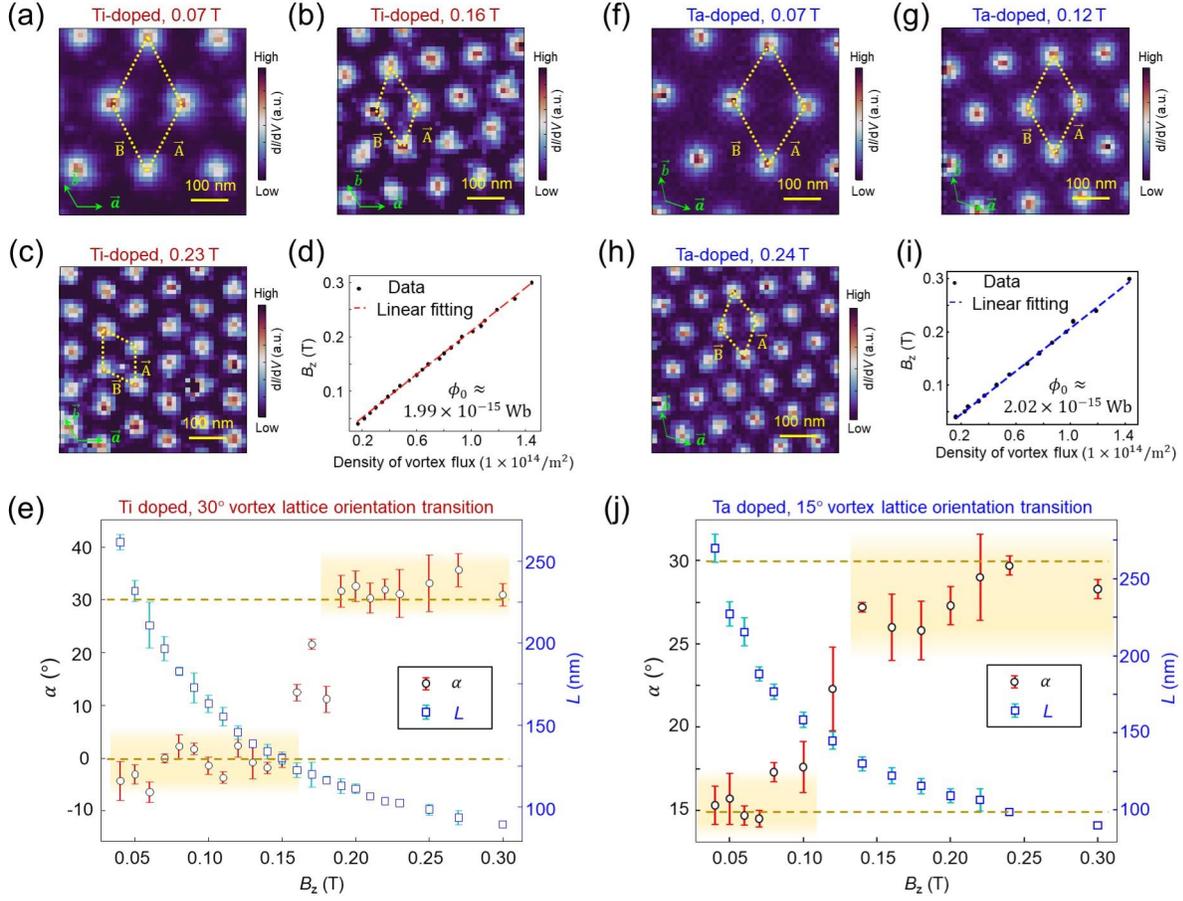

**Fig. 2.** The reorientation of vortex lattice in Ti-doped and Ta-doped $CsV_3Sb_5$. (a-c) Zero-energy d$I$/d$V$ maps (500 nm×500 nm) obtained at the same region of Ti-doped sample at 0.07 T, 0.16 T and 0.23 T, respectively, showing isotropic-shaped vortices and quasi-hexagonal vortex lattice. ($V$=−1 mV, $I$=500 pA, $V_{mod}$=0.1 mV). The crystal lattice is marked by green arrows with directional vector $\vec{a}$ and $\vec{b}$, and vortex lattice cell is highlighted by the orange rhombuses with directional vectors $\vec{A}$ and $\vec{B}$. (d) The magnetic field dependence of vortex density in Ti-doped $CsV_3Sb_5$. Red dotted line denotes the linear fitting of experimental data, giving the single magnetic flux quantum of $1.99\times10^{-15}$ Wb. (e) $B_z$-dependent vortex lattice orientation angle $\alpha$ and inter-vortex lattice $L$, showing a vortex lattice orientation transition with 30° rotation. (f-h) Zero-energy d$I$/d$V$ maps (500 nm×500 nm) obtained at the same region of Ta-doped sample at 0.07 T, 0.12 T and 0.24 T, respectively, showing isotropic-shaped vortices and quasi-hexagonal vortex lattice. ($V$=−1 mV, $I$=500 pA, $V_{mod}$=0.1 mV). The crystal lattice is marked by green arrows with directional vector $\vec{a}$ and $\vec{b}$, and vortex lattice cell is highlighted by the orange rhombuses with directional vectors $\vec{A}$ and $\vec{B}$. (i) The magnetic field dependence of vortex density in Ta-doped $CsV_3Sb_5$. Blue dotted line denotes the linear fitting of experimental data, giving the single magnetic flux quantum of $2.02\times10^{-15}$ Wb. (j) $B_z$-dependent $\alpha$ and $L$, showing a vortex lattice orientation transition with 15° rotation.

*3.3. Distinct spatial evolution of VBSs of Ti-doped and Ta-doped CsV$_3$Sb$_5$*

To gain a better understanding of the distinct superconductivity in the Ti-doped and Ta-doped samples, we investigated the VBSs inside the Abrikosov vortices. Due to the U-shaped gap pairing and nearly-zero conductance at the Fermi level, the spatial evolution of VBSs in the vortex of doped samples is clearer than that of pristine CsV$_3$Sb$_5$ with V-shaped gap pairing [31]. To compare the VBSs of two doped samples and avoid possible surface effect, we firstly focus on surface regions with dilute Cs adatoms (Fig. S3 online). At $B_z$=0.08 T, the size of the vortex core in the Ti-doped sample (Fig. 3a) is larger than that in the Ta-doped sample (Fig. 3d), indicating a longer coherent length of superconductivity in the former one (Fig. S4 online). The d$I$/d$V$ line-cut across the vortex core along distinct directions show isotropic features in both samples (see Supplementary Materials Fig. S5 and Fig. S6 online). Notably, in the Ti-doped sample, a sharp ZBCP emerges at the vortex core (Fig. 3b). Crossing the vortex, the ZBCP remains robust and non-split along a relatively-long distance (Fig. 3b, c), which displays an exotic Y-type spatial evolution feature. However, in the Ta-doped sample, the ZBCP only exist in the vortex center (Fig. 3e, f) and split right off as it moved away from the vortex center, displaying a typical X-type spatial evolution. To better identify the energy positions of the in-gap states, we plotted the negative second derivative of d$I$/d$V$ ($-$d$^3I$/d$V^3$) curves, which clearly show the Y-type VBSs spatial evolution for Ti-doped sample and X-type evolution for Ta-doped sample (right panel of Fig. 3c, f and Fig. S7 online). It is noteworthy that for less Ta doping content ($x$ =0.08), the VBSs still evolve as trivial X-type (Fig. S8d online), while for a higher Ti concentration ($x$ =0.09), the VBSs maintain the Y-type spatial evolution (Fig. S9b, c online).

The ZBCP spatial evolution is robust against magnetic field and surface regions (Supplementary text I and Fig. S10, S11, S12, S13 online), demonstrating that the distinct evolution of VBSs originates from the exotic bulk superconducting order parameter rather than previously reported surface-dependent results in pristine compound [31].

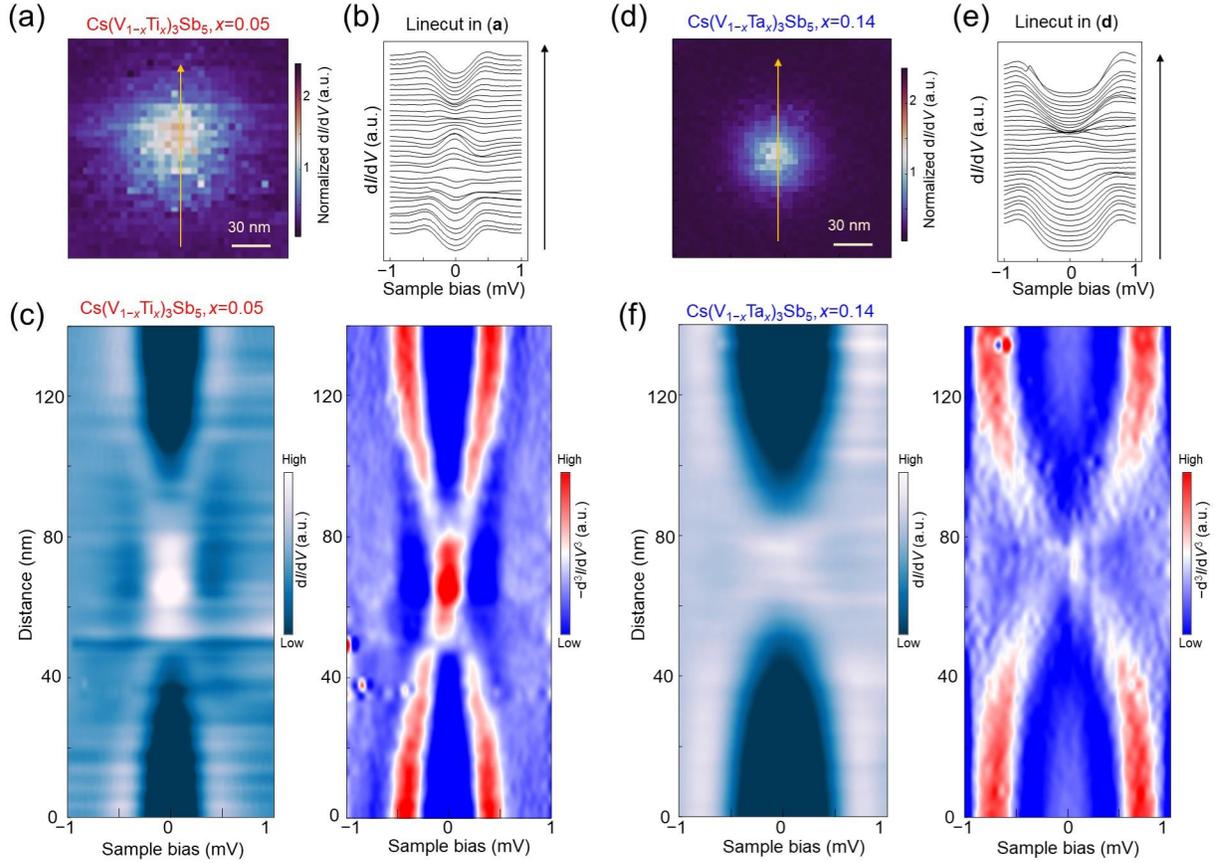

**Fig. 3.** Distinct spatial evolution of VBSs of Ti-doped and Ta-doped CsV$_3$Sb$_5$. (a) d$I$/d$V$ map at 0 meV showing a field-induced vortex at Sb surface of Ti-doped sample under a magnetic field of 0.08 T perpendicular to the surface ($V=-10$ mV, $I=1$ nA, $V_{mod}=0.1$ mV). (b) The d$I$/d$V$ spectra along the yellow arrow across the vortex in (a), showing a relatively-sharp zero-bias peak in the vortex core ($V=-10$ mV, $I=1$ nA, $V_{mod}=0.1$ mV). (c) Line-cut intensity plot and corresponding second-derivative plot of (b), showing Y-type spatial evolution of the VBSs. (d) d$I$/d$V$ map at 0 meV showing a field-induced vortex at Sb surface of Ta-doped sample under a magnetic field of 0.08 T perpendicular to the surface ($V=-10$ mV, $I=0.1$ nA, $V_{mod}=0.1$ mV). (e) The d$I$/d$V$ spectra along the yellow arrow across the vortex in (d), showing a relatively-broad zero-bias peak in the vortex core ($V=-10$ mV, $I=1$ nA, $V_{mod}=0.1$ mV). (f) Line-cut intensity plot and corresponding second-derivative plot of (e), showing X-type spatial evolution of the in-gap bound states inside vortex.

*3.4. Tunable superconductivity and VBSs in various chemically-doped CsV$_3$Sb$_5$ and the possible origin of the distinct VBSs*

To summarize the results for various chemically doped samples, we present the $-d^3I/dV^3$ spectra ($dI/dV$ spectra can be found in Fig. S14a online) obtained at the vortex core center in $Cs(V_{1-x}Ta_x)_3Sb_5$ ($x$ =0.14, 0.08), pristine $CsV_3Sb_5$ and $Cs(V_{1-x}Ti_x)_3Sb_5$ ($x$ =0.05, 0.09) at a magnetic field of 0.08 T (Fig. 4a). Interestingly, the ZBCPs are strong and sharp for Ti-doped samples, but weak and broad for Ta-doped and pristine samples (Fig. 4b). Additionally, the ZBCP of $x$=0.05 is sharper than that of $x$=0.09 for Ti-doped samples. The full width of the half maximum (FWHM) of the ZBCP crossing the vortex can be extracted from the $dI/dV$ spectra across the vortex core by Gauss fitting (Fig. S14b online). We define the spatial length of ZBCP before splitting to be the non-splitting decay length ($l$), representing the length over which the FWHM in Gauss fitting remain continuous without dramatic change. The FWHMs measured at $Cs(V_{1-x}Ti_x)_3Sb_5$ ($x$=0.05) are small (~0.36 meV) within a long non-splitting decay length ($l$~30 nm). In $Cs(V_{1-x}Ti_x)_3Sb_5$ ($x$ =0.09), the FWHMs are comparable to $x$=0.05 but with a shorter decay length ($l$~20 nm). In contrast, the measured FWHMs in pristine and Ta-doped $CsV_3Sb_5$ are large and can only be maintained over a short distance ($l$<5 nm). $Cs(V_{1-x}Ti_x)_3Sb_5$ ($x$=0.05) has the longest decay distance for the ZBCP, while $Cs(V_{1-x}Ta_x)_3Sb_5$ ($x$=0.14) has the shortest one.

Figure 4c illustrates a schematic superconducting phase diagram of chemically-doped $CsV_3Sb_5$ based on transport results [35] and observations from this study. In the non-doped or diluted-doped range, the material is in the multiband superconducting-I phase (SC-I) with an unconventional superconducting V-shaped gap. This phase coexists with long-range unidirectional $4a_0$ and bidirectional $2a_0$ CDWs, and the vortices at Sb surfaces host trivial CdGM states with a short decay length of ZBCP. Both Ta- and Ti-doped $CsV_3Sb_5$ exhibit a U-shaped paring gap that is concurrent with short-range stripe order at the surface (Fig. S15 online). However, the difference in vortex lattice orientation transition and spatial evolution of VBSs distinguishes the two superconducting phases. For Ti doping, the material is in the multiband superconducting-II phase (SC-II) with a U-shape pairing gap. The vortex lattice undergoes a 30º orientation transition, and the vortices at both Cs and Sb surfaces host non-trivial VBSs with unconventional Y-type spatial evolution. In contrast, for Ta doping, the material is in the multiband superconducting-III phase (SC-III), exhibiting a size-enhanced U-shaped gap. The vortex lattice undergoes 15º orientation transition, and the ZBCP in the vortex core of both Cs and Sb surfaces are weak and broad, behaving as a conventional X-type spatial evolution.

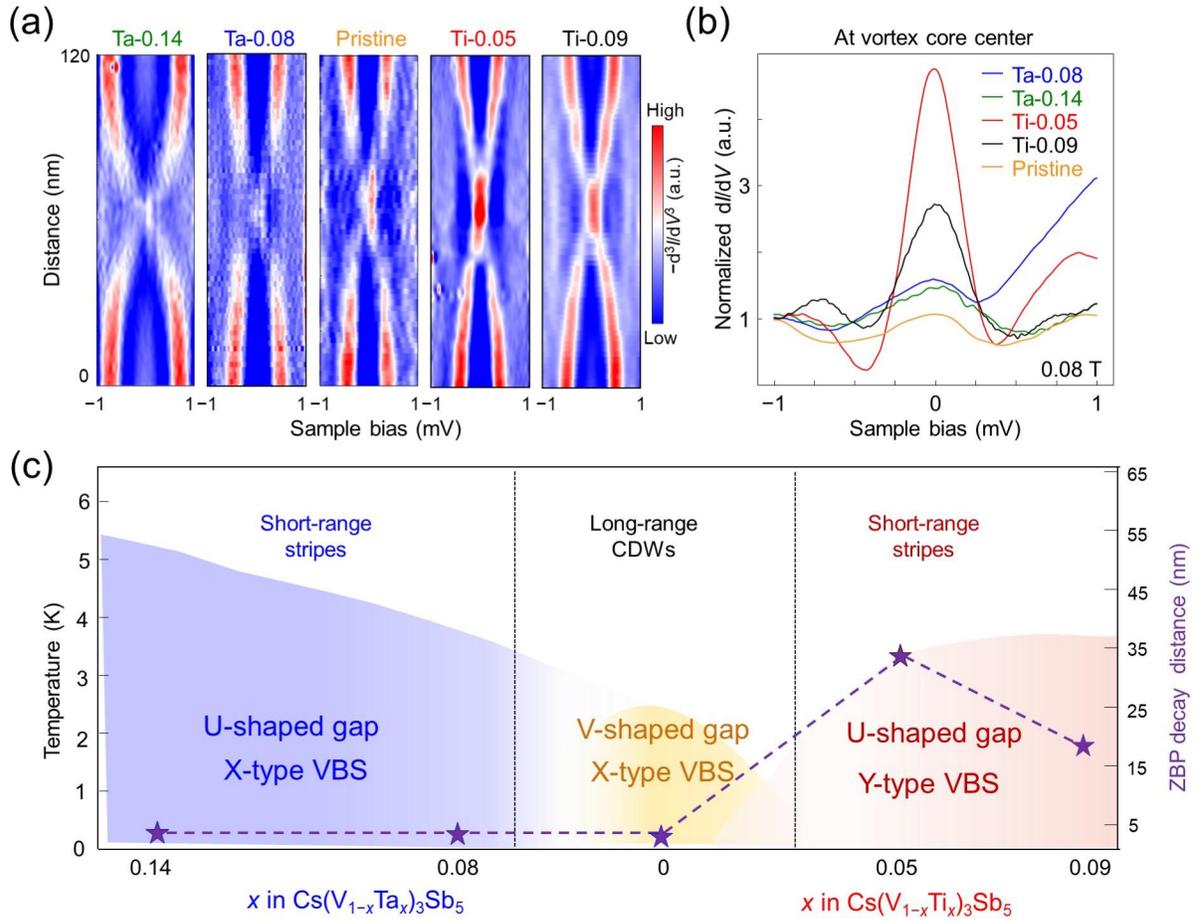

**Fig. 4.** Tunable superconductivity and VBSs in various chemically-doped $CsV_3Sb_5$ and the possible origin of the distinct VBSs. (a) $-d^3I/dV^3$ linecut across vortex core of $Cs(V_{1-x}Ta_x)_3Sb_5$ ($x$=0.14, 0.08), pristine $CsV_3Sb_5$ and $Cs(V_{1-x}Ti_x)_3Sb_5$ ($x$=0.05, 0.09) at a magnetic field of 0.08 T, respectively, showing the tunable VBSs in various chemically-doped $CsV_3Sb_5$. (b) Stack plot of the normalized $dI/dV$ spectra at vortex core of distinct doped sample, showing that the zero-bias peak for the 0.05-Ti doped sample is the sharpest. (c) Phase diagram for the Ti-doped and Ta-doped sample. Three distinct superconducting phases are identified. The undoped sample exhibits V-shaped superconducting gap paring and X-type spatial evolution of VBSs (I-phase) and the Ta doped sample shows U-shaped superconducting gap paring and X-type spatial evolution of VBSs (III-phase). In contrast, the Ti doped sample presents U-shaped gap pairing and Y-type spatial evolution of VBSs (II-phase).

The emergence of a non-split ZBCP over long distances in the Ti-doped $CsV_3Sb_5$ system suggests the presence of exotic physics. For a trivial vortex that doses not reach the quantum limit, the evolution of the VBSs has the relation [48] of $E_{VBS}(r) \sim \frac{\partial E_\mu}{\partial \mu} k_F r$, which corresponds to an X-type evolution. Therefore, the trivial nature of the vortex core in Ta-doped CsV$_3$Sb$_5$ is supported by the X-type spatial evolution of CdGM states. However, for a Y-type spatial evolution, there are various possibilities. It is commonly understood that a non-split ZBCP is a clear indication of the existence of zero-energy modes. In a nodeless superconductor, a vortex with zero-energy bound state typically requires an additional π phase to shift the $\frac{1}{2}\frac{\Delta^2}{E_F}$ of CdGM states to zero energy [5,9], or a special anisotropic Fermi surface [49]. Based on the similar Fermi surface structure between Ta- and Ti-doped CsV$_3$Sb$_5$ (Fig. S16c online) and the isotropy of the vortex (Fig. S5 and Fig. S6 online), the special anisotropic Fermi surface case can be excluded. Therefore, the Y-type spatial evolution of VBSs may indicates a topological vortex based on the Fu-Kane model [50] or the chiral superconductor [5,51,52]. There are various studies that have theoretically and experimentally attributed the non-split Y-type evolution to the MZM [8,50,53]. Thus, the first possible explanation for the Y-type VBSs evolution is the existence of MZM, considering the TSSs picture of the Fu-Kane model [14]. In pristine CsV$_3$Sb$_5$, several TSSs are predicted to lie above Fermi level, and Y-type VBSs evolution has been observed on the Cs surface due to the possible electron doping effect of TSSs [31]. However, it has been demonstrated that Ti substitution simply introduces a hole doping effect [35], which will push the predicted TSSs that originally sit above the Fermi surface further away. Therefore, the scenario for the MZM based on the Fu-Kane model may be excluded. Another possible scenario for topological vortices with chiral order parameter also lacks strong evidence.

We discuss another possible origin of the spatial evolution of VBSs beyond MZM scenario. Our model is founded on an isotropic Fermi surface and *s*-wave pairing, and we solve the Eilenberger equation to obtain the VBSs (Fig. S17 online). We have determined that the zero-energy peak's characteristic length is proportional to $\xi = v_F/\Delta_0$, where $v_F$ is the Fermi velocity and $\Delta_0$ is the size of the superconducting gap. By utilizing the experimentally extracted $\Delta_0$ and assuming Fermi velocity $v_F^{Ti} = 2.5 v_F^{Ta}$, we have qualitatively reproduced the Y-type evolution of Ti-doped sample (Fig. S17a online) and the X-type evolution of Ta-doped sample (Fig. S17b online). The crucial point to consider is why $v_F$ of Ti-doped sample is much greater than Ta-doped sample. To address this issue, we have calculated $v_F$ (Fig. S16 online) for both the Ti-doped and Ta-doped samples (k$_z$=π in Fig. S17 and k$_z$=0 in Fig. S16c online). The overall $v_F$ has not undergone significant alterations, with both the $v_F$ of Sb p$_z$ orbital (central red circle orbital in Fig. S17c, d online) at around 600 km/s and the $v_F$ of V *d* orbital at around 300 km/s (Fig. S17c, d online). Consequently, one plausible explanation is that the dissimilar orbital weighs in superconducting pairing could lead to the significant difference in $v_F$ between the Ta- and Ti-doped samples. More experimental evidences on the orbital selected Cooper pairing and origin of non-split zero-bias vortex core states in the chemically-doped $CsV_3Sb_5$ are required in feature works.

## 4. Summary

In conclusion, we observed tunable VBSs in two family of mutliband $CsV_3Sb_5$-derived kagome superconductors. The VBSs show X-type VBSs spatial evolution in Ta-doped $CsV_3Sb_5$ while non-split Y-type VBSs spatial evolution in Ti-doped ones. The Y-type VBSs evolution was previously believed to be the signature of MZM in a non-quantum-limit condition. We discuss an alternative explanation based on the possible orbital difference between these two $CsV_3Sb_5$-derived kagome superconductors. Our results provide a brand-new insight in the community of exploring MZMs not in a quantum-limit system. Additionally, MZMs could potentially emergent in $CsV_3Sb_5$-derived kagome system upon electron doping to push TSSs closing Fermi level, which will be an attracting project for future research.

# References

[1] Abrikosov AA. Nobel lecture: type-II superconductors and the vortex lattice. Rev Mod Phys 2004; 76: 975–9.

[2] Song SY, Hua C, Bell L, et al. Nematically templated vortex lattices in superconducting FeSe. Nano Lett 2023; 23:2822–30.

[3] Duan W, Chen K, Hong W, et al. Bamboo-like vortex chains confined in canals with suppressed superconductivity and standing waves of quasiparticles. Nano Lett 2022; 22:9450–6.

[4] Caroli C, De Gennes PG, Matricon J. Bound Fermion states on a vortex line in a type II superconductor. Phys Lett 1964; 9:307–9.

[5] Volovik GE. Fermion zero modes on vortices in chiral superconductors. Jetp Lett 1999; 70:609–14.

[6] Zhou J, Wang S-Z, Wu Y-J, et al. Topological mid-gap states of $p_x + ip_y$ topological superconductor with vortex square superlattice. Phys Lett A 2014; 378: 2576–81.

[7] Franz M, Tešanović Z. Self-consistent electronic structure of a $d_{x^2-y^2}$ and a $d_{x^2-y^2} + id_{xy}$ Vortex. Phys Rev Lett 1998; 80: 4763–6.

[8] Kawakami T, Hu X. Evolution of density of states and a spin-resolved checkerboard-type pattern associated with the majorana bound state. Phys Rev Lett 2015; 115: 177001.

[9] Kong L, Zhu S, Papaj M, et al. Half-integer level shift of vortex bound states in an iron-based superconductor. Nat Phys 2019; 15: 1181–7.

[10] Wang Y, MacDonald AH. Mixed-state quasiparticle spectrum for $d$-wave superconductors. Phys Rev B 1995; 52: R3876–9.

[11] Gazdić T, Maggio-Aprile I, Gu G, et al. Wang-MacDonald $d$-wave vortex cores observed in heavily overdoped $Bi_2Sr_2CaCu_2O_{8+\delta}$. Phys Rev X 2021; 11: 031040.

[12] Nayak C, Simon SH, Stern A, et al. Non-Abelian anyons and topological quantum computation. Rev Mod Phys 2008; 80: 1083–159.

[13] Kitaev AYu. Fault-tolerant quantum computation by anyons. Ann Phys 2003; 303: 2–30.

[14] Fu L, Kane CL. Superconducting proximity effect and Majorana Fermions at the surface of a topological insulator. Phys Rev Lett 2008; 100: 096407.

[15] Chen X, Chen M, Duan W, et al. Robust zero energy modes on superconducting bismuth islands deposited on Fe(Te,Se). Nano Lett 2020; 20: 2965–72.

[16] Huxley A, Rodière P, Paul DMcK, et al. Realignment of the flux-line lattice by a change in the symmetry of superconductivity in $UPt_3$. Nature 2000; 406: 160–4.

[17] Cubitt R, Eskildsen MR, Dewhurst CD, et al. Effects of two-band superconductivity on the flux-line lattice in magnesium diboride. Phys Rev Lett 2003; 91: 047002.

[18] Fan P, Chen H, Zhou X, et al. Nanoscale manipulation of wrinkle-pinned vortices in iron-based superconductors. Nano Lett 2023; 23: 4541–7.

[19] M. Suzuki K, Inoue K, Miranović P, et al. Generic first-order orientation transition of vortex lattices in type II superconductors. J Phys Soc Jpn 2010; 79: 013702.

[20] Ortiz BR, Teicher SML, Hu Y, et al. $CsV_3Sb_5$: a $Z_2$ topological kagome metal with a superconducting ground state. Phys Rev Lett 2020; 125: 247002.

[21] Nie L, Sun K, Ma W, et al. Charge-density-wave-driven electronic nematicity in a kagome superconductor. Nature 2022; 604: 59–64.

[22] Li H, Zhao H, Ortiz BR, et al. Unidirectional coherent quasiparticles in the high-temperature rotational symmetry broken phase of $AV_3Sb_5$ kagome superconductors. Nat Phys 2023; 19: 637–43.

[23] Xiang Y, Li Q, Li Y, et al. Twofold symmetry of c-axis resistivity in topological kagome superconductor $CsV_3Sb_5$ with in-plane rotating magnetic field. Nat Commun 2021; 12: 6727.

[24] Chen H, Yang H, Hu B, et al. Roton pair density wave in a strong-coupling kagome superconductor. Nature 2021; 599: 222–8.
[25] Hu B, Ye Y, Huang Z, et al. Robustness of the unidirectional stripe order in the kagome superconductor $CsV_3Sb_5$. Chin Phys B 2022; 31: 058102.
[26] Zhao H, Li H, Ortiz BR, et al. Cascade of correlated electron states in the kagome superconductor $CsV_3Sb_5$. Nature 2021; 599: 216–21.
[27] Mielke C, Das D, Yin J-X, et al. Time-reversal symmetry-breaking charge order in a kagome superconductor. Nature 2022; 602: 245–50.
[28] Yu L, Wang C, Zhang Y, et al. Evidence of a hidden flux phase in the topological kagome metal $CsV_3Sb_5$ ArXiv: 2107. 10714, 2021.
[29] Ni S, Ma S, Zhang Y, et al. Anisotropic superconducting properties of kagome metal $CsV_3Sb_5$. Chin Phys Lett 2021; 38: 057403.
[30] Xu H-S, Yan Y-J, Yin R, et al. Multiband superconductivity with sign-preserving order parameter in kagome superconductor $CsV_3Sb_5$. Phys Rev Lett 2021; 127: 187004.
[31] Liang Z, Hou X, Zhang F, et al. Three-dimensional charge density wave and surface-dependent vortex-core states in a kagome superconductor $CsV_3Sb_5$. Phys Rev X 2021; 11: 031026.
[32] Chen KY, Wang NN, Yin QW, et al. Double superconducting dome and triple enhancement of $T_c$ in the kagome superconductor $CsV_3Sb_5$ under high pressure. Phys Rev Lett 2021; 126: 247001.
[33] Yu F, Zhu X, Wen X, et al. Pressure-induced dimensional crossover in a kagome superconductor. Phys Rev Lett 2022; 128: 077001.
[34] Yu FH, Ma DH, Zhuo WZ, et al. Unusual competition of superconductivity and charge-density-wave state in a compressed topological kagome metal. Nat Commun 2021; 12: 3645.
[35] Yang H, Huang Z, Zhang Y, et al. Titanium doped kagome superconductor $CsV_{3-x}Ti_xSb_5$ and two distinct phases. Sci Bull 2022; 67: 2176–85.
[36] Oey YM, Ortiz BR, Kaboudvand F, et al. Fermi level tuning and double-dome superconductivity in the kagome metal $CsV_3Sb_{5-x}Sn_x$. Phys Rev Mater 2022; 6: L041801.
[37] Ortiz BR, Gomes LC, Morey JR, et al. New kagome prototype materials: discovery of $KV_3Sb_5$, $RbV_3Sb_5$, and $CsV_3Sb_5$. Phys Rev Mater 2019; 3: 094407.
[38] Zhong Y, Liu J, Wu X, et al. Nodeless electron pairing in $CsV_3Sb_5$-derived kagome superconductors. Nature 2023; 617: 488–92.
[39] Kresse G, Furthmüller J. Efficiency of ab-initio total energy calculations for metals and semiconductors using a plane-wave basis set. Comput Mater Sci 1996; 6: 15–50.
[40] Perdew JP, Burke K, Ernzerhof M. Generalized gradient approximation made simple. Phys Rev Lett 1996; 77: 3865–8.
[41] Grimme S, Antony J, Ehrlich S, et al. A consistent and accurate ab initio parametrization of density functional dispersion correction (DFT-D) for the 94 elements H-Pu. J Chem Phys 2010; 132: 154104.
[42] Luo Y, Han Y, Liu J, et al. A unique van Hove singularity in kagome superconductor $CsV_{3-x}Ta_xSb_5$ with enhanced superconductivity. Nat Commun 2023; 14: 3819.
[43] Hanaguri T, Kitagawa K, Matsubayashi K, et al. Scanning tunneling microscopy/spectroscopy of vortices in LiFeAs. Phys Rev B 2012; 85: 214505.
[44] Zehetmayer M. How the vortex lattice of a superconductor becomes disordered: a study by scanning tunneling spectroscopy. Sci Rep 2015; 5: 9244.
[45] Avers KE, Kuhn SJ, Leishman AWD, et al. Reversible ordering and disordering of the vortex lattice in $UPt_3$. Phys Rev B 2022; 105: 184512.
[46] Zhitomirsky ME, Dao V-H. Ginzburg-Landau theory of vortices in a multigap superconductor. Phys Rev B 2004; 69: 054508.

[47] Das P, Rastovski C, O'Brien TR, et al. Observation of well-ordered metastable vortex lattice phases in superconducting $MgB_2$ using small-angle neutron scattering. Phys Rev Lett 2012; 108: 167001.
[48] Gygi F, Schlüter M. Self-consistent electronic structure of a vortex line in a type-II superconductor. Phys Rev B 1991; 43: 7609–21.
[49] Kim H, Nagai Y, Rózsa L, et al. Anisotropic non-split zero-energy vortex bound states in a conventional superconductor. Appl Phys Rev 2021; 8: 031417.
[50] Xu J-P, Wang M-X, Liu ZL, et al. Experimental detection of a Majorana mode in the core of a magnetic vortex inside a topological insulator-superconductor $Bi_2Te_3$/$NbSe_2$ heterostructure. Phys Rev Lett 2015; 114: 017001.
[51] Lee D, Schnyder AP. Structure of vortex-bound states in spin-singlet chiral superconductors. Phys Rev B 2016; 93: 064522.
[52] Sato M, Takahashi Y, Fujimoto S. Non-Abelian topological orders and Majorana fermions in spin-singlet superconductors. Phys Rev B 2010; 82: 134521.
[53] Lv Y-F, Wang W-L, Zhang Y-M, et al. Experimental signature of topological superconductivity and Majorana zero modes on β-$Bi_2Pd$ thin films. Sci Bull 2017; 62: 852–6.